\documentclass[pdflatex,sn-mathphys-num]{sn-jnl}

\usepackage[pdftex, dvipsnames]{xcolor}
\usepackage[T1]{fontenc} 
\usepackage{chemformula} 
\usepackage[separate-uncertainty=true, 
            multi-part-units=single]{siunitx} 
\usepackage{fix-cm}

\DeclareSIUnit{\sample}{S}
\DeclareSIUnit\bit{b}
\usepackage{makecell,tabularx}
\setcellgapes{3pt}
\usepackage{braket}
\setcitestyle{numbers,square}

\usepackage{xargs}
\usepackage[disable,colorinlistoftodos,prependcaption,textsize=tiny]{todonotes}
\newcommandx{\unsure}[2][1=]{\todo[linecolor=red,backgroundcolor=red!25,bordercolor=red,#1]{#2}}
\newcommandx{\change}[2][1=]{\todo[linecolor=blue,backgroundcolor=blue!25,bordercolor=blue,#1]{#2}}
\newcommandx{\FIXME}[1][1=]{\todo[linecolor=blue,backgroundcolor=blue!25,bordercolor=blue,#1]{FIXME}}
\newcommandx{\info}[2][1=]{\todo[linecolor=OliveGreen,backgroundcolor=OliveGreen!25,bordercolor=OliveGreen,#1]{#2}}
\newcommandx{\improvement}[2][1=]{\todo[linecolor=Plum,backgroundcolor=Plum!25,bordercolor=Plum,#1]{#2}}
\newcommandx{\addref}[1][1=]{\todo[linecolor=red,backgroundcolor=red!25,bordercolor=red,#1]{Add reference}}

\newcommand\gtwo[2][]{\ifthenelse{\isempty{#2}}{$g^{(2)}(0)$}{\ifthenelse{\isempty{#1}}{$g^{(2)}(0)=#2$}{$g^{(2)}(0)=#2\pm#1$}}}

\begin{document}
\title{Entanglement-verified time distribution in a metropolitan network}

\author[1]{Mohammed K. Alqedra} 
\equalcont{These authors contributed equally to this work.}
\author[1]{Samuel Gyger}
\equalcont{These authors contributed equally to this work.}
\author[1]{Katharina D. Zeuner}
\author[1]{Thomas Lettner}
\author[2]{Mattias Hammar}
\author[3]{Gemma Vall Llosera}
\author*[1]{Val Zwiller}\email{zwiller@kth.se}

\affil[1]{\orgdiv{Department of Applied Physics}, \orgname{KTH Royal Institute of Technology}, \street{Roslagstullsbacken 21}, \postcode{106 91}, \orgaddress{\city{Stockholm}, \country{Sweden}}}
\affil[2]{\orgdiv{Department of Electrical Engineering}, \orgname{KTH Royal Institute of Technology}, \street{Electrum 229} \postcode{164 40} \city{Kista}, \country{Sweden}}
\affil[3]{\orgname{Ericsson AB}, \orgaddress{\street{Torshamnsgatan 21}, \city{Stockholm}, \postcode{16440}, \country{Sweden}}}

\abstract{The precise synchronization of distant clocks is a fundamental requirement for a wide range of applications. 
Here, we experimentally demonstrate a novel approach of quantum clock synchronization utilizing entangled and correlated photon pairs generated by a quantum dot at telecom wavelength. 
By distributing these entangled photons through a metropolitan fiber network in the Stockholm area and measuring the remote correlations, we achieve a synchronization accuracy of tens of picoseconds by leveraging the tight time correlation between the entangled photons. 
We show that our synchronization scheme is secure against spoofing attacks by performing a remote quantum state tomography to verify the origin of the entangled photons. 
We measured a distributed maximum entanglement fidelity of \num{0.817 \pm 0.040} to the $\ket{\Phi^+}$ Bell state and a concurrence of \num{0.660 \pm 0.086}. 
These results highlight the potential of quantum dot-generated entangled pairs as a shared resource for secure time synchronization and quantum key distribution in real-world quantum networks.
}

\maketitle

Geographically distributed quantum entanglement is a fundamental resource for a wide range of quantum applications, including quantum communication \cite{Kimble2008Jun, Wehner2018Oct}, distributed quantum computing \cite{Cirac1999Jun} and distributed quantum sensing \cite{Guo2020Mar}. 
A crucial requirement for the realization of these applications is the highly precise, secure, and robust time synchronization of remote clocks. 
In quantum key distribution (QKD), precise and secure time synchronization is essential for accurately associating detection events with their corresponding photons between the communicating parties, thereby enabling the successful generation of a shared secret key \cite{Basset2021Mar}. 
In addition, synchronized clocks can help mitigate background noise by temporal filtering of incoming signals expected within a predefined time window, thereby enhancing the signal-to-noise ratio in applications such as quantum sensing \cite{Malia2022Dec, Zhang2021Jul}.

In classical networks, clock synchronization is achieved through hierarchical systems that rely on accurate time sources to distribute time across the network \cite{Mills2002Aug}. 
At the top of the hierarchy are highly accurate clocks, such as those based on Global Navigation Satellite Systems (GNSS) like Global Positioning System (GPS), which provide precise time signals using atomic clocks on satellites.
These time signals are received by ground-based devices, such as GNSS receivers, which act as the primary reference clocks for synchronization systems.
The accurate time distribution across the network is subsequently achieved through synchronization protocols such as Network Time Protocol (NTP) and Precision Time Protocol (PTP). 
In NTP, time flows through a tiered structure, starting from Stratum-0 clocks (e.g., GNSS receivers or atomic clocks) to Stratum-1 servers, which directly synchronize to the primary time source. These servers then pass the time to lower levels (Stratum-2, Stratum-3, etc.), progressively synchronizing devices.
Although high synchronization accuracy can be achieved with classical synchronization schemes, such as White Rabbit Precision Time Protocol (WRPTP), which can reach sub-nanosecond precision under optimized conditions \cite{Lipinski2011}, these methods are still vulnerable to certain security limitations. 
One significant concern is their susceptibility to spoofing attacks, where an adversary can manipulate synchronization signals to disrupt or falsify the timing information since the origin of the signal cannot be verified. 

During the last two decades, several quantum clock synchronization methods based on entangled and correlated photons have been proposed and implemented to improve the accuracy and security of the time distribution \cite{Jozsa2000Aug,Giovannetti2001Jul, Giovannetti2002Jan}. 
These methods leverage the tight time correlations between correlated photon pairs to enhance the synchronization accuracy and enable a more precise time distribution. 
In addition, the security of the time distribution can be enhanced by measuring non-local quantum properties of the shared photon pairs, verifying their origin and providing protection against spoofing attacks. Typically, previously demonstrated protocols utilized correlated photons generated through spontaneous parametric down-conversion (SPDC).

In the two-way time transfer over fiber scheme \cite{Hou2019Aug,Lee2019Mar,Quan2022Mar}, an SPDC source is used in each of the nodes to be synchronized. 
One photon from each correlated pair is detected locally, whereas the other is detected at the remote node after propagating through the communication channel. 
Assuming symmetric propagation delays, the position of the correlation peak in each node marks the one-way propagation time, enabling synchronization. However, these SPDC-based schemes relied only on timing correlations, without experimentally verifying quantum entanglement. This limitation restricts their capability to authenticate photon origin, thus making them potentially vulnerable to spoofing or intercept attacks and limiting their security guarantees.
Clock synchronization has also been demonstrated using bidirectionally propagating photons generated by a single SPDC source \cite{Lee2022May}. 
Despite the high accuracy of clock synchronization achieved by these methods, they remain vulnerable to asymmetric delay attacks \cite{Lee2019Sep}. 
In addition, the probabilistic nature of SPDC sources used in clock synchronization schemes introduces a trade-off between the efficiency and security of the time transfer. 

Here, we demonstrate an absolute time synchronization between two clock nodes at remote locations in the Stockholm metropolitan area using bidirectionally propagating entangled and correlated photon pairs generated by a triggered quantum dot source. 
We verify the authenticity of the synchronization method by adding delay lines of a known length measured independently, namely \SI{4.48}{\nano\second} and \SI{0.30}{\nano\second}, and measure the absolute offsets between the two synchronized clocks on the order of tens of picoseconds. 
In addition, we perform remote quantum state tomography to verify the origin of the shared photons, thus providing protection against spoofing attacks. 
The use of a telecom-wavelength quantum dot source further supports integration with fiber-based quantum communication protocols, offering a platform for unified timing and key distribution in future quantum networks.


\section{Setup}

\begin{figure*}[htb]

\includegraphics[width=\textwidth]{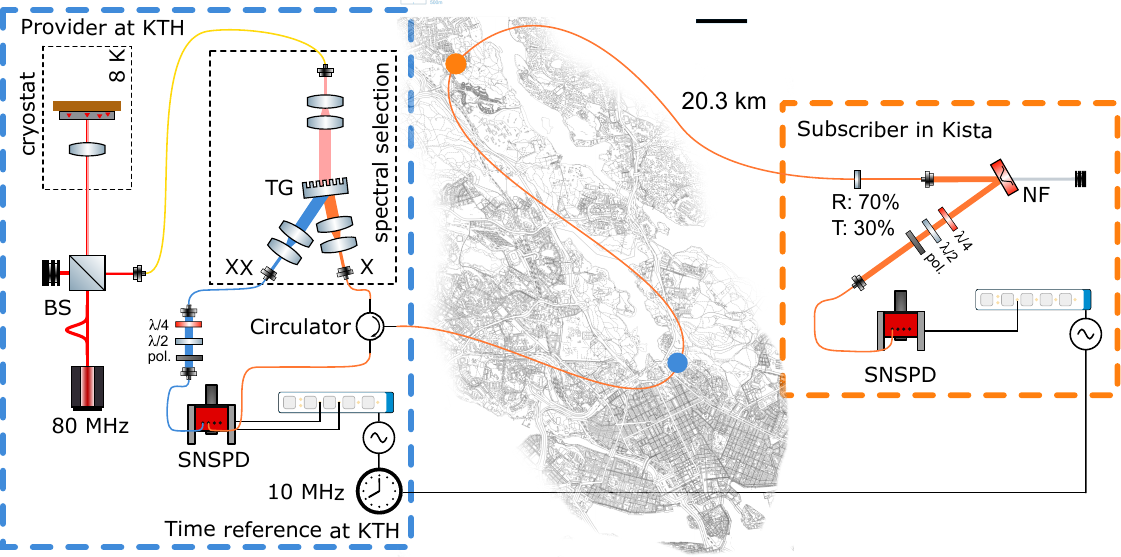}
\caption{\label{fig:fig1-setup} 
{\textbf{Metropolitan fiber link in Stockholm region for entanglement-verified time synchronization.} Metropolitan fiber linking the time reference node in downtown Stockholm, marked by the blue circle, to a subscriber in Kista, marked by the orange circle. Entangled photon pairs at $\approx\SI{1550}{\nano\meter}$ were generated using an InAs/GaAs quantum dot cooled down to $\approx\SI{8}{\kelvin}$ and hosted at the same node as the time reference. An 80 MHz source at 1470 nm was used to excite the quantum dot to the p-shell. A transmission grating (TG) was used to filter the bi-exciton and the exciton. The bi-exciton photons were detected at the local timing reference node. The exciton photons were combined with the path of two reference lasers used to create a polarization reference for polarization stabilization across the link (not shown in the figure). At the subscriber node, an in-line partial reflector was used to reflect \SI{70}{\percent} of the exciton back to the time reference node, where it was separated by a circulator and detected locally. The transmitted \SI{30}{\percent} of the exciton was filtered by a notch filter at the subscriber and detected by an SNSPD. For entanglement verification, a set of quarter waveplate ($\lambda/4$), half waveplate ($\lambda/2$) and a polarizer (pol.) were used to to measure 16 different polarization bases, which were used to reconstruct the density matrix at each time bin in the cascade. The scale bar is \SI{1}{\kilo\meter}.}
Map Data by OpenStreetMap Contributors. Map adapted with permission from Oliver O'Brien.
}
\end{figure*}

The experimental setup used for the entanglement distribution and clock synchronization is shown in Figure~\ref{fig:fig1-setup}. The provider of polarization entangled single photon pairs is located at the KTH Royal Institute of Technology in downtown Stockholm and connected through a \SI{20}{\kilo\meter} long fiber to a distant subscriber in Kista, a nearby industrial center. 
The pairs of entangled photons are generated by a self-assembled InAs/GaAs quantum dot at a wavelength of $\approx\SI{1550}{\nano\meter}$. 
A detailed description of the sample can be found in Ref. \cite{Zeuner2021Aug}.

The source was hosted within a cryostat at $\approx\SI{8}{\kelvin}$ and was coupled to a single-mode fiber using a high numerical aperture (NA) cryogenic objective (Attocube LT-APO/NIR NA=\num{0.8}). 
An \SI{80}{\mega\hertz} pulsed laser at \SI{1470}{\nano\meter} was used to excite the quantum dot into the p-shell. 
A detailed characterization of the quantum dot emission properties is presented in the supplementary material. 
The entangled photon pairs were generated via a biexciton (XX) and exciton (X) cascade. 
The biexciton was filtered using a reflective notch filter and sent to the local timing reference node. 
The exciton was combined with the path of two reference lasers (not shown in the figure), that were used to create a polarization reference across the link. 
The light was then fiber-coupled and transmitted through a polarization stabilized \SI{20}{\kilo\meter} public fiber connecting to the subscriber in Kista.
At the subscriber node, an electrically driven fiber polarization controller was used to control the output polarization before the quantum signal gets separated from the two reference lasers using a reflective and a transmitting notch filter. 
The correlation and polarization measurements were temporarily interleaved.
Both nodes have a polarization measurement module that includes a quarter waveplate, half waveplate and a polarizer before being detected by superconducting nanowire single photon detectors (SNSPD) and digitized with a time-to-digital converter (qutools quTAG).
The clock frequencies in the two nodes were locked to a \SI{10}{\mega\hertz} reference source transmitted via a second fiber paired with the main fiber, thereby sharing environmental influences. 
Time distribution is accomplished by synchronizing the frequency at both nodes and determining the absolute time offset between the two clocks.

\section{Results}
\subsection{Clock Synchronization}
To implement the timing synchronization protocol, \SI{70}{\percent} of the light at the subscriber node in Kista was reflected back, while the remaining photons were filtered by a reflective notch-filter to reject optical noise due to cross-talk from other signals within the network. After filtering, the photons were measured locally, detected by an SNSPD and timestamped by a time-to-digital converter. The reflected photons were separated using a circulator and detected at the master clock in provider location, after being filtered from network noise. The time stamps of the time reference were provided to the subscriber and can be combined to evaluate two correlations between the subscriber and the time reference. The correlations can then be used to get the propagation time of the photons in the fiber and the offset between the local and the master clocks. Given the low count rates, the cross-correlations were measured by acquiring coincidences over a duration of \num{10}-\num{20}~hours for each measurements.

\begin{figure*}[htb]
\centering
\includegraphics[width=\columnwidth]{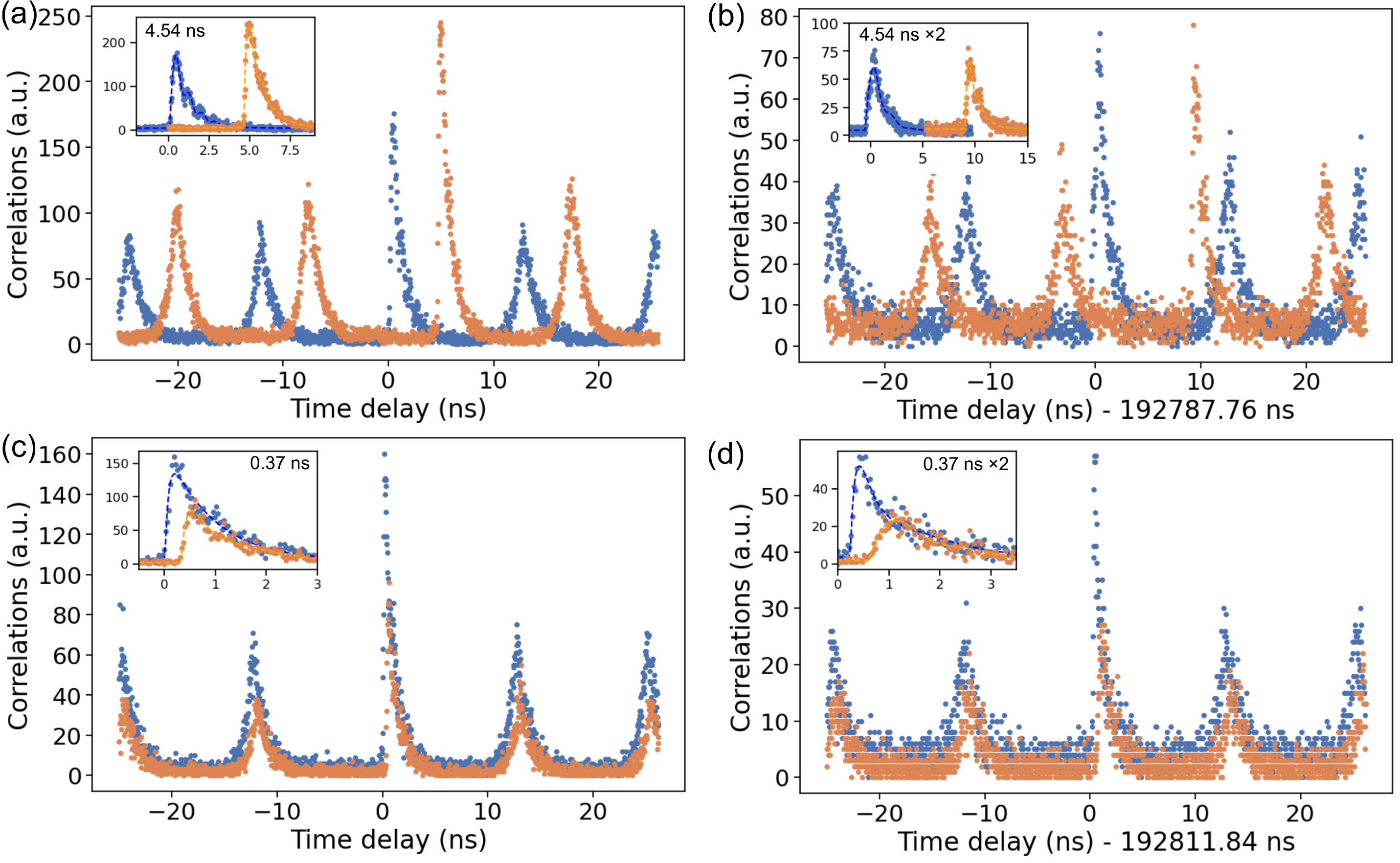}
\caption{\label{fig:fig3-time-synchronisation}
\textbf{Clock synchronization.} (a) and (c) Cross-correlation measurements between biexciton photons detected at the master clock node and the \SI{30}{\percent} of exciton photons detected at the subscriber node in Kista, before (blue) and after (orange) adding known fiber delays. Insets provide zoomed-in views around the correlation peaks, clearly showing the shift caused by the added delays.
(b) and (d) A cross-correlation measurement between biexciton photons and returned exciton photons, both detected at the master clock node, before (blue) and after (orange) adding the fiber delays. Insets show close-up views of the correlation peaks. 
This measurement reveals the a round-trip time of \SI{192.78776}{\micro\second}, which is subtracted from the x-axis. Combining both measurements is used to find the absolute offset between the two clocks.
The added fiber delays extracted from the measurements in (a) and (b) is \SI{4.54}{\nano\second}, closely matching the independently measured value of \SI{4.48}{\nano\second} obtained through time-of-flight measurement. 
In (c) and (d), we measured a delay of \SI{0.37}{\nano\second} compared to \SI{0.30}{\nano\second} measured independently by time of flight. 
The extra \SI{24.09}{\nano\second} subtracted from the x-axis of (d) with respect to (b) corresponds to the fiber length to connect the delay line used for the \SI{0.30}{\nano\second} measurement. 
The insertion of the \SI{0.30}{\nano\second} delay line introduced approximately \SI{1}{dB} of additional loss.
The orange data in (c) and (d) with the \SI{0.30}{\nano\second} delay was integrated over \num{10}~hours, while all other correlation measurements were integrated over \num{20}~hours each.
}
\end{figure*} 

The first correlation was measured between the biexciton photons directly detected at the master clock and the transmitted part of the exciton photons detected at the subscriber clock in Kista.
This correlation measurement reveals the offset between the subscriber and the master clocks.
The second correlation was measured between the directly detected biexciton photons at the master clock and the round-trip returned photons that were partially reflected back from the subscriber node and detected at the master clock node. 
This second correlation reveals the propagation time of the photons in the fiber. 
These two measurements provide the absolute offset between the two clocks and allow for high-precision time distribution. 
We measure the offset between the clocks on the order of \num{10}s of picoseconds, and correct for the round-trip time of \SI{192.78776}{\micro\second} utilizing the reflected photons. 
To verify the authenticity of the method we added fiber delays of \SI{4.48}{\nano\second} and \SI{0.30}{\nano\second}, both segments measured separately by time-of-flight, while keeping the two clocks locked to the same \SI{10}{\mega\hertz} clock. 
The cross-correlations measured between the biexciton photons and the exciton photons at the subscriber node are shown in Figure~\ref{fig:fig3-time-synchronisation}(a) for \SI{4.48}{\nano\second} fiber delay and (c) for \SI{0.30}{\nano\second} delay, before (blue) and after (orange) adding the delays. 
Figure~\ref{fig:fig3-time-synchronisation}(b) for \SI{4.48}{\nano\second} and (d) for \SI{0.30}{\nano\second} show the cross-correlations between the biexciton photons and the round-trip exciton photons, both detected at the master clock node. 

For the \SI{4.48}{\nano\second} fiber delay, we observed an extension of the round-trip time by $2\times\SI{4.54}{\nano\second}$, and an absolute offset between the master and subscriber clock of \num{918251.66950915298}~seconds, compared to \num{918251.66950914837}~seconds before adding the extra fiber delay. 
To ensure consistency, this measurement was performed twice with two different quantum dots, both yielding similar results, thereby verifying the robustness and reproducibility of the synchronization protocol. 
The results from the second quantum dot are discussed in the supplementary material. 
It should be noted that, in addition to the optical and electrical path difference between the two clocks, this absolute clock offset also includes the intrinsic time offset between the two clocks, arising from their different initialization times.
This intrinsic offset was identified independently using an optical signal at the start of the measurement, to reduce the search space for the correlation peak, on a $\approx \SI{100}{\milli\second}$ accuracy.

For the added \SI{0.30}{\nano\second} delay line, we observed an extension of the round-trip time by $2\times\SI{0.37}{\nano\second}$, and an absolute offset between the master and subscriber clocks of \num{22218.91375984974}~seconds, compared to \num{22218.91375984933}~seconds before adding the delay line.

It should be noted that the precision of the time distribution depends on how accurately the cross-correlation peak can be identified. In this work, we determine the maximum position of the cross-correlation peak by fitting it with a theoretical model capturing the exponential rise and modulated exponential decay due to the fine-structure splitting of the biexciton-exciton emission cascade. A detailed description of the fitting function and its parameters is provided in the supplementary material. The fit of each measurement is shown in the insets shown in Figure \ref{fig:fig3-time-synchronisation}.

Due to the relatively low emission rate of the quantum dot used in this work, identifying the exact position of the maximum in the cross-correlation peak involves some uncertainty, despite the relatively long integration times. Specifically, the count rates at different nodes were approximately as follows: biexciton photons detected locally at the master clock node at \num{12} kcps; exciton photons detected remotely at the subscriber node in Kista at \num{500} cps; and partially reflected exciton photons detected back at the master clock node at \num{250} cps. The uncertainty is higher for correlation measurements involving returned exciton photons, as they experience twice the fiber losses and potentially greater exposure to environmental fluctuations before being detected at the master clock node. Using a brighter quantum dot source would reduce this uncertainty and improve synchronization precision.

\begin{figure*}[htb]
\includegraphics[width=\columnwidth]{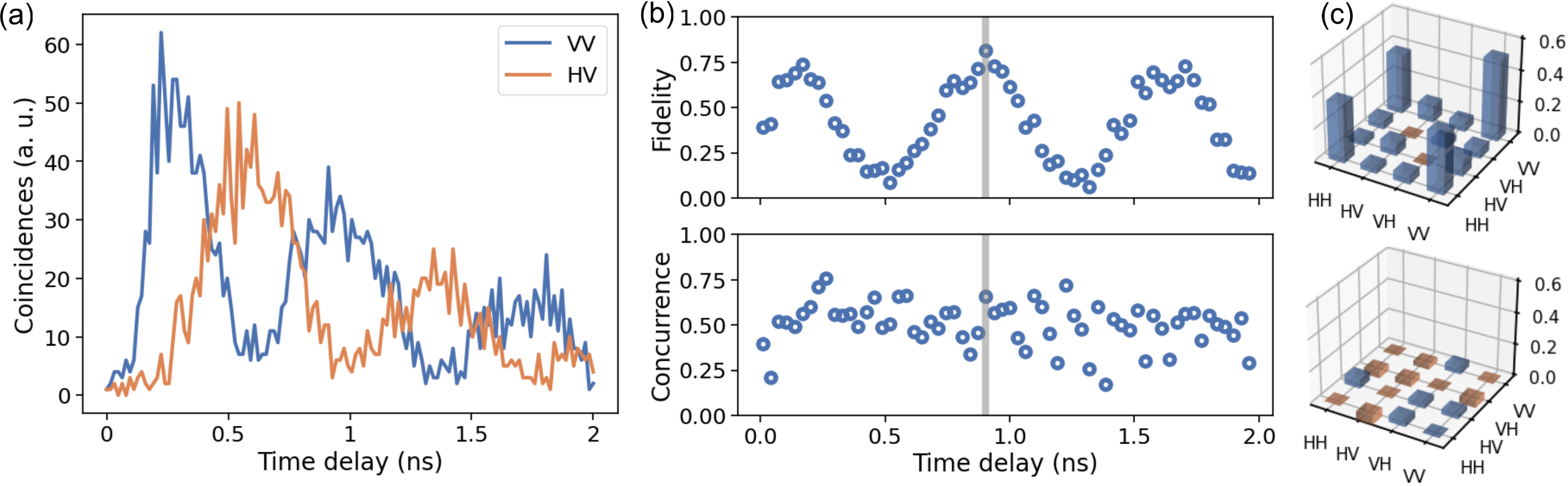}
\caption{\label{fig:fig2-entanglement_distribution}
\textbf{Entanglement distribution over \SI{20}{\kilo\meter} of metropolitan network.} (a) Measurement of the coincidence at different time delays for the VV and HV polarization projections. (b) The time evolved fidelity of the two-photon states to the $\ket{\Phi^+}$ state (top panel), and concurrence (bottom panel), extracted at different time delays. The oscillation in the coincidence and fidelity plots are due to the fine-structure splitting. (c) The reconstructed density matrix at the time-bin with maximum fidelity to the $\ket{\Phi^+}$ state.
}
\end{figure*}

The intrinsic timing resolution of our detection, limited primarily by the detection jitter of the SNSPDs and time-to-digital converters, is approximately \SI{100}{\pico\second}. While accumulating coincidence counts over extended integration periods (10–20 hours) reduces statistical uncertainties in locating correlation peaks, uncertainties from environmental effects, such as temperature-induced fiber-length variations, must also be considered. This has previously been measured on the same deployed fiber, showing that thermal fluctuations could introduce drifts of up to \SI{1}{\nano\second} over a \num{20}~hours period \cite{Staffas2023Feb}. However, our synchronization protocol significantly mitigates such environmental drifts, since the fiber carrying the quantum dot photons and the paired fiber carrying the classical frequency reference experience the same temperature-induced length variations. Moreover, the long integration periods inherently average out slow frequency drifts. Any significant residual drift or fiber asymmetry would manifest as broadening or systematic shifts in the correlation peaks, neither of which was significant in our data. Therefore, we conservatively estimate our practical synchronization uncertainty to be approximately less than 100 ps, primarily limited by the stability of the reference clock and the detection jitter.

\subsection{Entanglement Verification} 

To evaluate the shared polarization entanglement between the two nodes, we performed remote quantum state tomography on the two-photon state generated by the biexciton-exciton cascade. We measured a total of 16 time tagged data corresponding to 16 different projective measurements on the joint biexciton-exciton polarization state [see supplementary material]. 
We use the extensible time tag analyzing (ETA) software to analyze the data and extract the correlation histograms \cite{Lin2021Aug}. 
The correlation histograms corresponding to the HH and HV projections are shown in Figure~\ref{fig:fig2-entanglement_distribution} (a). 
The observed oscillations in the coincidences are due to the fine-structure splitting, which causes the emitted photon state to temporally oscillate between the two Bell states $\ket{\Phi^+} = \frac{1}{\sqrt{2}} (\ket{HH} + \ket{VV})$ and $\Phi^- = \frac{1}{\sqrt{2}} (\ket{HH} - \ket{VV})$. 
The high time-resolution of our SNSPDs, along with precise frequency synchronization between the two clocks [see supplementary material], enables us to resolve this evolution. 
The fidelity of the measured two-photon state to the $\ket{\Phi^+}$ Bell state and the concurrence are shown in Figure~\ref{fig:fig2-entanglement_distribution} (b), with the reconstructed density matrix at the time-bin with the maximum fidelity shown in Figure~\ref{fig:fig2-entanglement_distribution} (c).
We observe a maximum fidelity of \num{0.817 \pm 0.040} to the $\Phi^+$ state for the time delay of \SI{896}{\pico\second} highlighted in gray in Figure~\ref{fig:fig2-entanglement_distribution}(b), and a concurrence of \num{0.660 \pm 0.086} at the same time bin. 
For comparison, a local tomography measurement on both photons at the provider location revealed a maximum fidelity of \num{0.907 \pm 0.028} and a maximum concurrence of \num{0.886 \pm 0.044}. 
A detailed discussion of the local tomography measurement is presented in the supplementary material. The degradation of the entanglement quality when distributed between the two nodes is likely due to the limitation of the active polarization stabilization used throughout the measurement. Nevertheless, the relatively high entanglement fidelity observed between the two nodes verifies the origin of the photons used for clock synchronization.

The distributed entangled photon pairs can be used as a resource for entanglement-based QKD protocols \cite{Basset2021Mar,Yang2024Jul}. 
The use of quantum dot sources for QKD offers several advantages, such as the low multiphoton emission rate, which can be further minimized using two-photon excitation techniques \cite{Schweickert2018Feb}, on-demand emission at telecom wavelengths, which is crucial for long-distance communications \cite{Yu2023Dec}. 
High emission rates have been realized from quantum dots around \SI{900}{\nano\meter} \cite{Tomm2021Apr,Chen2018Jul}. 
Improvements are being explored to enhance the brightness at telecom wavelengths by utilizing techniques such as Purcell enhancement \cite{Yu2023Dec, Vajner2024Feb}. 
Using entangled photon pairs for both clock synchronization and entanglement-based QKD could reduce the overall hardware requirements since part of the emitted photons can be used for synchronization while the rest are utilized for key distribution. 
This dual-purpose use of entangled photons streamlines the system, enhancing efficiency and reducing the complexity of the hardware setup required for secure communications and precise timekeeping.

\section{Discussion and Conclusion}
In this work, we demonstrated a quantum verified clock synchronization scheme between two remote locations separated by a \SI{20}{\kilo\meter} long fiber based on triggered entangled and time-correlated photons, addressing several security vulnerabilities present in timing protocols. Our approach enables the distribution of entangled photons from a central node to remote subscribers, ensuring that the origin of the synchronization signal can be verified and potential tampering within the fiber network can be detected.

A fundamental concern in secure synchronization is the risk of interception and spoofing attacks. An adversary could attempt to inject false timing signals into the network. To mitigate this, we verify the authenticity of the received photons through remote quantum state tomography performed at both the provider and subscriber nodes. This verification ensures that only photons originating from the trusted entangled source contribute to the synchronization process.

Another inherent challenge in any time synchronization protocol is the assumption of symmetric propagation delays. While free-space transmission naturally maintains symmetry, fiber-based networks introduce vulnerabilities due to the potential for asymmetric delays. For instance, an adversary could introduce a circulator and delay line to passively modify the effective propagation time in one direction. Our experiment does not directly provide absolute immunity to such attacks; however, under practical constraints, it offers a reasonable degree of resilience. Assuming symmetric loss in the fiber link, the insertion of nontrivial components—such as circulators or reflectors—may introduce measurable asymmetries in timing correlations or link transmission, particularly if these elements are not perfectly lossless or polarization-preserving. While this assumption does not preclude sophisticated adversaries \cite{Lee2019Sep}, it raises the technical barrier for undetectable manipulation and motivates the use of entanglement verification as an additional layer of protection.

Furthermore, entangled photons cannot be amplified without destroying their entanglement \cite{Wootters1982Oct}. This enables us to verify the origin of the photons and therefore to precisely measure the losses. By using a deterministic high-purity single-photon source, we further reduce the uncertainty associated with coherent-state attacks and ensure that no injected photons interfere with the synchronization process, which relies exclusively on genuine entangled pairs.
Our quantum synchronization protocol complements a classical frequency distribution scheme, where a \SI{10}{\mega\hertz} rubidium-derived reference is transmitted via RF-over-fiber to maintain long-term clock stability [see supplementary material]. Although effective, classical synchronization alone remains vulnerable to undetected spoofing or asymmetric delays. By independently reconstructing high-fidelity entanglement correlations using these synchronized clocks, our method rigorously certifies the absence of significant unnoticed manipulation or relative frequency drift during measurements. This hybrid quantum-classical approach thus significantly strengthens security and trust in metropolitan-scale timing networks.

A key limitation in long-distance entanglement distribution is the mismatch between the optimal fiber transmission window at \SI{1550}{\nano\meter} and the typical emission wavelength of many quantum dot sources around \SI{900}{\nano\meter} \cite{Tomm2021Apr,Chen2018Jul}. This issue can be addressed through quantum frequency conversion \cite{Zaske2012Oct, Morrison2023Jun,Yang2024Jul, Zahidy2024Jan}, or by utilizing quantum dots engineered to emit at telecom wavelengths \cite{Zeuner2021Aug}. Additionally, improving the brightness of quantum emitters would enable shorter measurement times and enhance synchronization precision. The use of two-photon resonant excitation techniques could further improve the fidelity of the entangled photon pairs.

Overall, our results demonstrate highly precise and entanglement-verified clock synchronization across a \SI{20}{\kilo\meter} metropolitan fiber network. By leveraging the deterministic generation of entangled and correlated photon pairs from a quantum dot, our scheme achieves practical synchronization accuracy on the order of tens of picoseconds, limited by detection jitter and stability of frequency reference. We verify the authenticity of the synchronization signals via remote quantum state tomography, confirming a distributed maximum entanglement fidelity of \num{0.817 \pm 0.040} to the $\ket{\Phi^+}$ Bell state. This demonstration offers a potential reduction in overall hardware complexity by enabling the dual-purpose utilization of a single quantum resource for both synchronization and entanglement-based quantum key distribution. Our work establishes a viable approach to high-precision, secure quantum clock synchronization for quantum networks.

\section{Acknowledgments}
The authors thank Lucas Schweickert, Sandra Bensoussan, Liselott Ekemar, Noah Gläser, and Martin Carlnäs for their valuable support, as well as Carl Reuterskiöld Hedlund for assistance with sample preparation. We thank STOKAB for support with the fiber infrastructure.

\bibliography{sn-bibliography}
\end{document}


\title{Supplementary Material: Entanglement-verified time distribution in a metropolitan network}

\author[1]{Mohammed K. Alqedra} 
\equalcont{These authors contributed equally to this work.}
\author[1]{Samuel Gyger}
\equalcont{These authors contributed equally to this work.}
\author[1]{Katharina D. Zeuner}
\author[1]{Thomas Lettner}
\author[2]{Mattias Hammar}
\author[2]{Gemma Vall Llosera}
\author*[1]{Val Zwiller}\email{zwiller@kth.se}

\affil[1]{\orgdiv{Department of Applied Physics}, \orgname{KTH Royal Institute of Technology}, \street{Roslagstullsbacken 21}, \postcode{106 91}, \orgaddress{\city{Stockholm}, \country{Sweden}}}
\affil[2]{\orgdiv{Department of Electrical Engineering}, \orgname{KTH Royal Institute of Technology}, \street{Electrum 229} \postcode{164 40} \city{Kista}, \country{Sweden}}
\affil[3]{\orgname{Ericsson AB}, \orgaddress{\street{Torshamnsgatan 21}, \city{Stockholm}, \postcode{16440}, \country{Sweden}}}

\maketitle
\section{Theory of Correlation-based Clock Synchronization}

To mathematically formulate the synchronization procedure \cite{Bahder2004Nov}, we define the set of photon arrival times recorded by the time-to-digital converters within a well-defined time window, $\Delta$, at the provider and the subscriber clocks as {$t_{i,A}$} and {$t_{j,B}$}, where $i = 1, ..., N_A$, and $j = 1, ..., N_B$. Here, $N_A$ and $N_B$ represent the total number of timestamps recorded at the provider and subscriber nodes, respectively. 

The cross correlation, $C(\tau)$, which counts the number of times a photon detected at the provider node is coincident with a photon detected at subscriber node within the time window centered around $\tau$, can be written as:
\begin{equation}
C(\tau) = \sum_{i=1}^{N_A} \sum_{j=1}^{N_B} \Pi_\Delta(t_{i,A} - t_{j,B} - \tau),
\end{equation}

where $\Pi_\Delta(x)$ is the rectangular function defined as:
\begin{equation}
\Pi_\Delta(x) =
\begin{cases} 
1 & \text{if } |x| \leq \Delta/2 \\
0 & \text{otherwise}
\end{cases}
\end{equation}

We can write the normalized cross-correlation function $G^{(2)}(\tau)$ as:
\begin{equation}
G^{(2)}(\tau) = \frac{1}{\sqrt{N_{A} N_{B}}} \sum_{i=1}^{N_{A}} \sum_{j=1}^{N_{B}} \Pi_\Delta(t_{i,A} - t_{j,B} - \tau),
\end{equation}

Assuming that the frequency of both clocks is synchronized, only time stamps with $i=j$ will contribute to the correlations. The synchronization problem thus simplifies to finding the offset time $\delta\tau = t_{i,A} - t_{i,B}$. By adding this offset delay to the subscriber's clock B, it becomes synchronized with the provider's clock A. 

\section{Cross-correlation Peak Fitting Model}
In this work, we identify the maximum of the cross correlation peak by fitting it to the following function:
\begin{equation}
\begin{aligned}
f(t) &= A \left[1 - e^{\left(\frac{-(t - t_0)}{\tau_{\text{rise}}}\right)}\right] \cdot e^{\left(\frac{-(t - t_0)}{\tau_{\text{decay}}}\right)} \cdot \left[1 + B \cos(\omega (t - t_0) + \phi)\right] + C, \\
\end{aligned}
\label{Eq:cascade_fitting}
\end{equation}

\noindent where $A$ is the amplitude of the exponential rise and decay, $B$ is the amplitude of the oscillations, $\tau_{\text{rise}}$ is the time constant for the rise, $\tau_{\text{decay}}$ is the time constant for the decay, $\omega$ is the angular frequency of the oscillations due to the fine-structure splitting, $\phi$ is the phase of the oscillations, $C$ is the background level, and $t_0$ is the start time of the cascade rise. The first term in Equation \ref{Eq:cascade_fitting} refers to the rise time of the cascade. The second and the third terms are the decay of the cascade with the oscillation caused by the fine-structure splitting. 

\section{Spectral characterization of the quantum dot}

\begin{figure*}[htb]
\includegraphics[width=1\columnwidth]{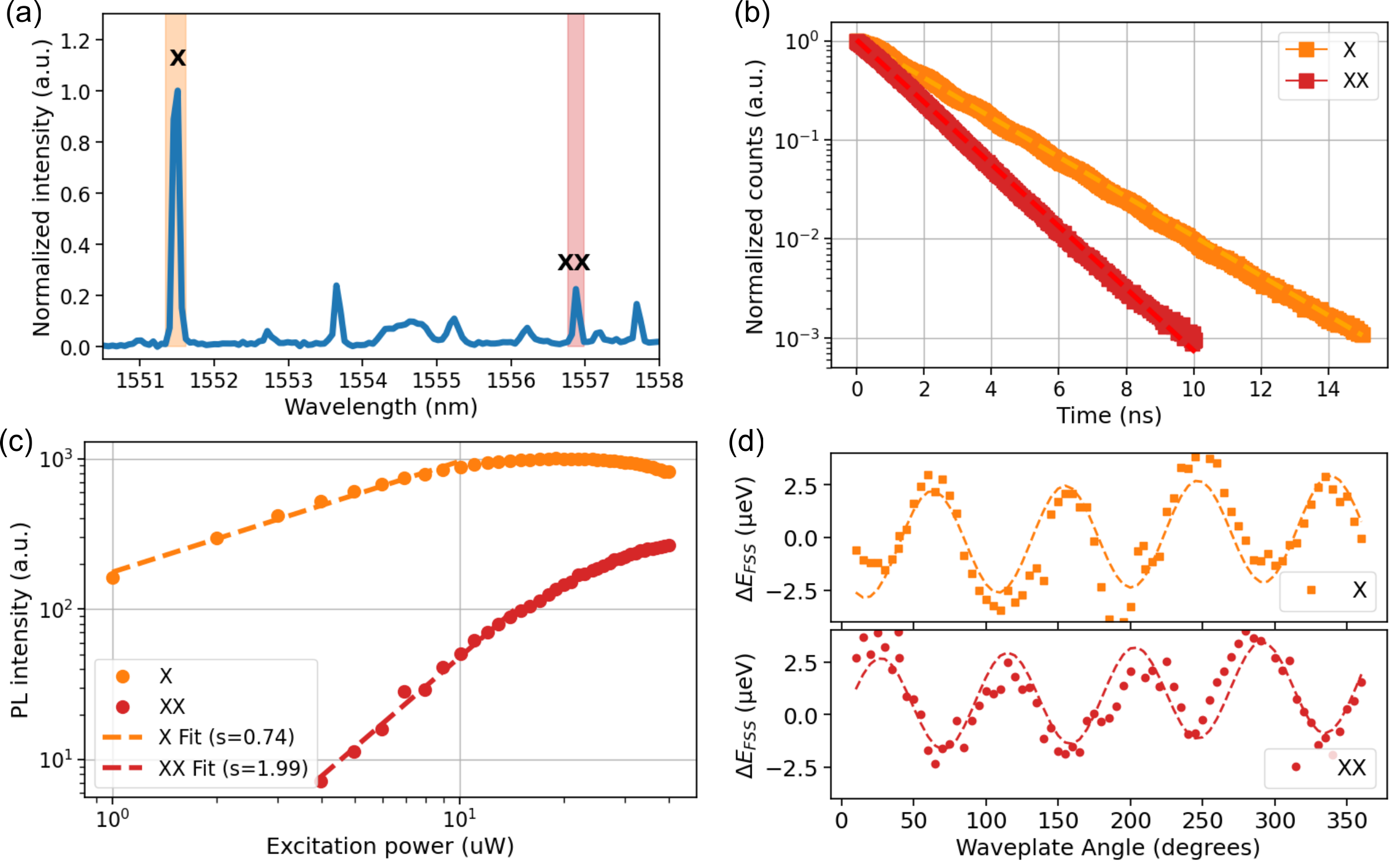}
\caption{\label{smfig:fig1-setup} 
\textbf{Spectral characterization of the quantum dot emission.}
(a) Emission spectra at $\sim \SI{7}{\kelvin}$ under p shell excitation. The exciton and biexciton are highlighted in orange and red, respectively. 
(b) The lifetime of the exction and the biexciton. 
(c) Power dependence of the exiton and biexciton. 
(d) Polarization-dependent measurement for both of X (orange) and XX (red) to determine the fine-structure splitting.
}
\end{figure*}

\noindent The quantum dot investigated in this study was characterized spectrally and temporally to confirm its suitability for high-fidelity entangled photon pair generation. 
Figure~\ref{smfig:fig1-setup} (a) shows the emission spectra of the quantum dot at $\sim\SI{7}{\kelvin}$ when excited p-shell at \SI{1470}{\nano\meter}. 
The exciton and bi-exciton transition are highlighted in orange and red, respectively. 
Furthermore, we examine the radiative lifetime of the exciton and the biexciton when excited to the p shell. 
The measured radiative decays were fitted to a single exponential, as shown in Figure~\ref{smfig:fig1-setup}(b). 
From the fit, we obtain a radiative lifetimes of \SI{2.17 \pm 0.007}{\nano\second} for the exciton (orange), and \SI{1.38 \pm 0.002}{\nano\second} for the biexciton (red). 
It should be noted that the measured exciton radiative lifetime does not directly correspond to its intrinsic radiative lifetime, as it reflects the entire cascaded emission process. 
The actual radiative lifetime of the exciton, determined from cross-correlation measurements, is \SI{1.14(1)}{\nano\second}.

Figure~\ref{smfig:fig1-setup}(c) presents the power-dependent photoluminescence measurements. 
The exciton exhibits an expected close to linear power dependence with a fitted slope of \num{0.74}, whereas the biexciton follows the expected quadratic power dependence with a slope of \num{1.99}. 

The fine structure splitting (FSS) of the exciton was characterized using polarization-dependent photoluminescence spectroscopy. 
A quarter-wave plate and a polarizing beam splitter were used to resolve the polarization components as the waveplate was rotated from \num{0} to \num{360}~degrees. 
The resulting oscillatory behavior, shown in Figure~\ref{smfig:fig1-setup}(d), reveals an average FSS of \SI{4.71}{\eV}.
This small splitting indicates a small anisotropic exchange interaction, which is favorable for high-fidelity entangled photon generation.

\section{Purity of Exciton and Biexciton}

\begin{figure*}
\includegraphics[width=\columnwidth]{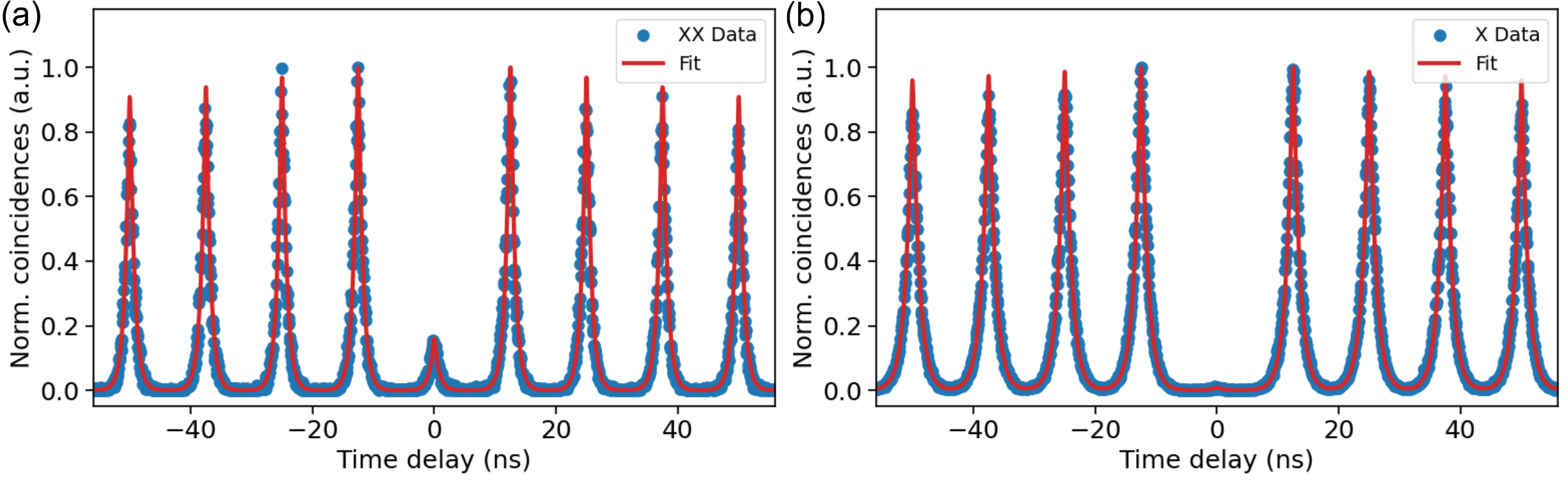}
\caption{\label{smfig:fig2-purity} 
\textbf{Autocorrelation measurement under p shell excitation.} 
(a) The exciton autocorrelation, and 
(b) the biexciton autocorrelation. 
The dashed line is a lorentzian fit for each peak, with the shadded areas showing the extracted full width at half maximum (FWHM) from the fit.
}
\end{figure*}

We measured the second-order auto-correlation function $g^{(2)}(t)$ for both the exciton and biexciton under p-shell excitation at \SI{1470}{\nano\meter}. 
This was achieved by separating the exciton and biexciton photons using a transmission spectrometer, sending the collected photons into a fiber-based beam splitter, and detecting the split counts using two SNSPDs.
The measured auto-correlation functions are shown in Figure~\ref{smfig:fig2-purity} (a) for the exciton and Figure~\ref{smfig:fig2-purity} (b) for the biexciton. 
The autocorrelation data was fitted using the model in Ref. \cite{Vajner2024Feb}, which corresponds to a sum of two-sided exponential decays while accounting for the blinking effect. The fit function was further convolved with a \SI{50}{\pico\second} Gaussian to account for the instrument response function (IRF). From the fit, we obtained $g_x^{(2)}(0)$=\num{0.010(7)} for the exciton, and $g_{xx}^{(2)}(0)=\num{0.15(4)}$ for the biexciton,. 
The higher multi-photon probability for the biexciton is likely due to spectral overlap with emission from nearby quantum dots.

\section{Local quantum state tomography}
\begin{figure*}[htb]
\includegraphics[width=\columnwidth]{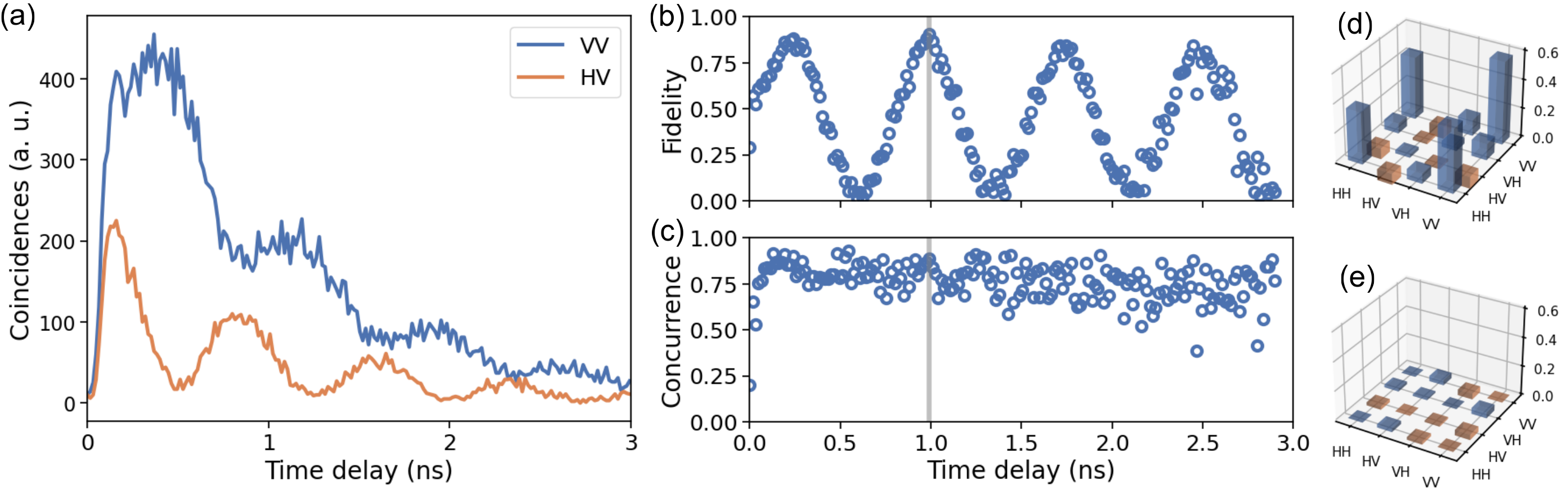}
\caption{\label{smfig:fig2-entanglement} 
\textbf{Local quantum state tomography of the entangled photons.} 
(a) Coincidence measurement of the VV and HV polarization projections. 
(b) The time evolved fidelity to the $\ket{\Phi^+}$ state and (c) concurrence of the two-photon states at different time delays. 
The oscillation in the coincidences and fidelity plots are due to the fine-structure splitting. (d) The real and (e) the imaginary part of the reconstructed density matrix at the highlighted time-bin in (b) and (c).
}
\end{figure*}

We characterize the quality of the entangled state at the provider location by performing sixteen coincidence measurements between the two entangled photons in sixteen polarization projections. To reconstruct the density matrices from the measured polarization bases, we use the maximum likelihood estimation algorithm used by Fognini et al.~\cite{afognini2025Jan}. 
In order to compensate for systematic polarization rotation and birefringence in our setup, we apply a virtual waveplate rotation with $\theta = \num{0.4545}~\text{rad}$ and $\phi = \num{-0.6763}~\text{rad}$, to transform the calculated raw density matrix to the correct HV basis, which is used for the polarization analysis.
Figure~\ref{smfig:fig2-entanglement} (a) shows the coincidence counts in two bases, namely, HH and HV, with the first letter refer to the measured polarization of the exciton photon, and the second letter to the polarization of the biexciton photon. 
The modulation in the coincidence counts is due to the fine-structure in the quantum dot transitions.
The sixteen coincidence measurements were used to reconstruct the density matrix at each of the 16 ps time bins throughout the cascade. 
The time evolved fidelity to the $\Phi^+$ bell state and the concurrence are shown in Figure~\ref{smfig:fig2-entanglement} (b) and (c), respectively. 
We measure a maximum fidelity of \num{0.907 \pm 0.028} to the $\Phi^+$ state and a concurrence of \num{0.886 \pm 0.044} for \SI{992}{\pico\second} time delay, marked in gray in Figure~\ref{smfig:fig2-entanglement} (b) and (c).
The real and imaginary parts of density matrix for this time delay with the maximum fidelity are shown in Figure~\ref{smfig:fig2-entanglement} (d) and (e), respectively.

\section{Frequency synchronization over fiber}
We use two time to digital converters (TDC) (qutools quTAG) located in both laboratories with a specified timing jitter of \SI{7}{\pico\second}.
Both are synchronized with an external \SI{10}{\mega\hertz} reference. A rubidium clock (SRS FS725) as a frequency reference is used in the KTH location.
The reference signal is transferred over a second fiber along the signal fiber at \SI{1310}{\nano\meter} using RF over fiber electronics (RFOptic Programmable 2.5GHz RFo) 
This allows the two TDCs to remain in sync throughout several days of measurement time.
We expect that the machine in the remote laboratory drifts with respect to time-of-flight distance changes induced by temperature changes. 
Utilizing the fiber pair allows  the clock to evolve along with the clock at the reference place, and keep the effective communication distance between the two clocks constant.
More information utilizing a comparable time tagger (Swabian TimeTagger Ultra) can be found in an Application Note \cite{Gyger.Glaser.ea:2023}.

\section{Entanglement verified time distribution using a second QD}

\begin{figure*}[htb]
\centering
\includegraphics[width=0.7\columnwidth]{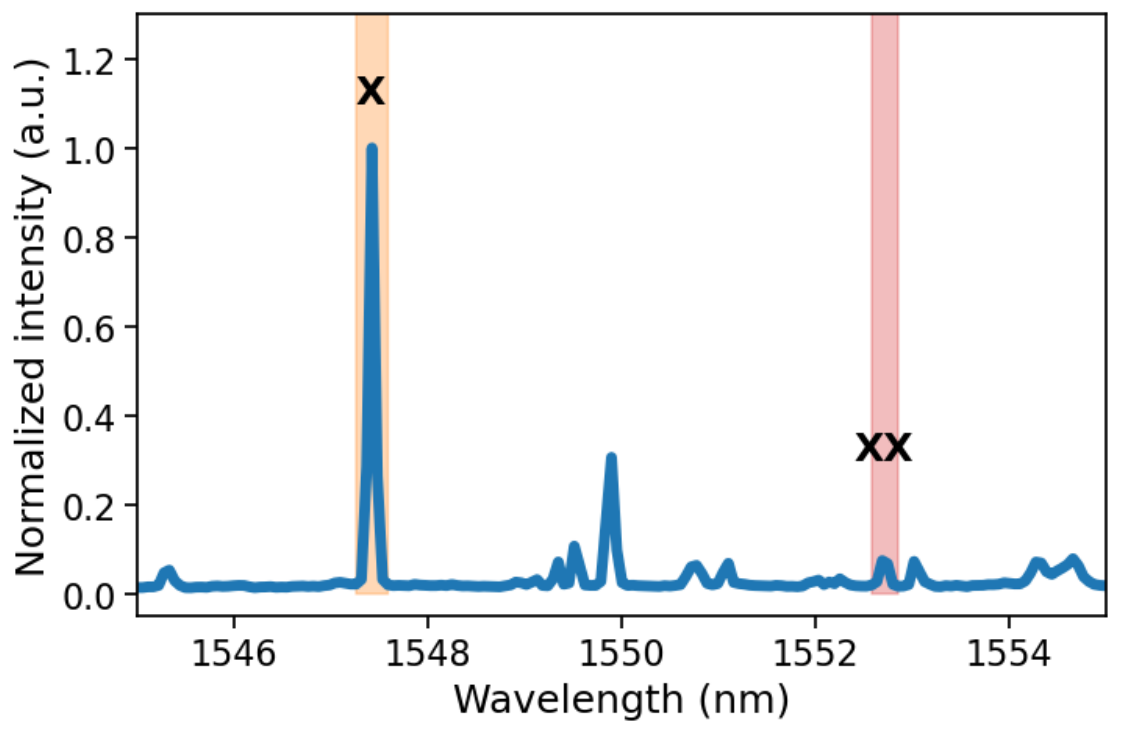}
\caption{\label{smfig:fig4-spectra_QD2} 
\textbf{The emission spectra of QD2 at about \SI{7}{\kelvin} when excited p-shell at \SI{1470}{\nano\meter}.}
The exciton and bi-exciton transition are highlighted in orange and red, respectively.
}
\end{figure*}

\begin{figure*}[htb]
\includegraphics[width=\columnwidth]{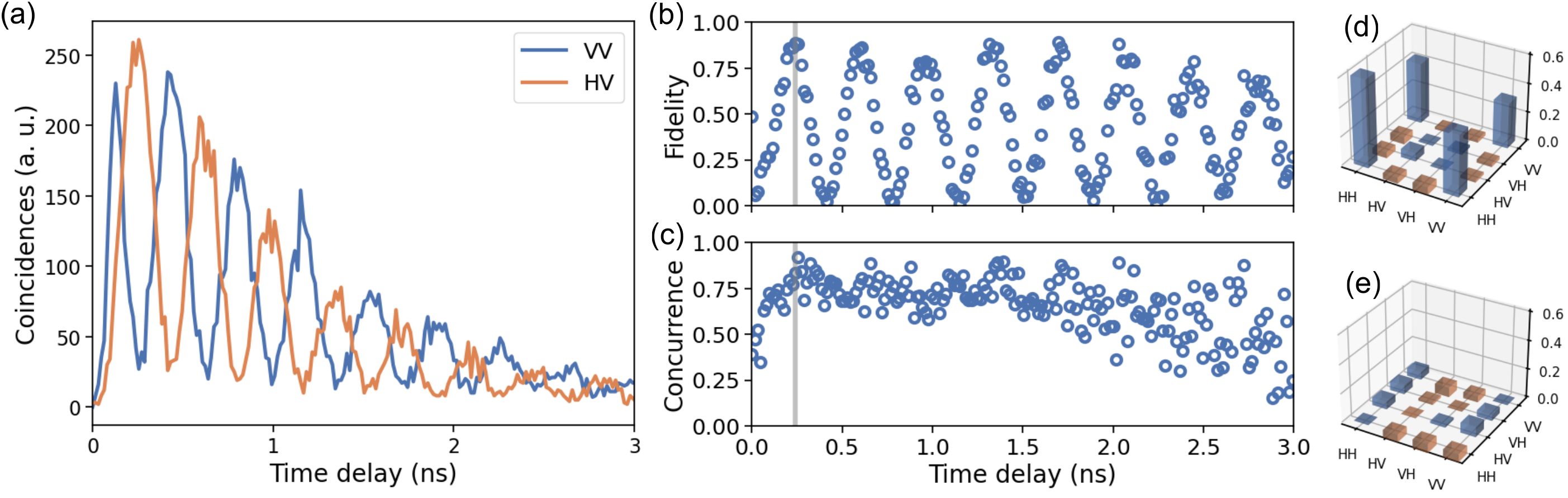}
\caption{\label{smfig:fig5-local_ent_QD2} 
\textbf{Local quantum state tomography for the entangled pairs generated by QD2.} 
(a) Measured coincidences for the VV and HV polarization projections. 
(b) Time evolved fidelity to the $\Phi^+$ target state as a function of time delay. 
(c) Concurrence as a function of time delay. 
(d) Real and (e) Imaginary parts of the reconstructed density matrix at the time highlighted time bin in (b) and (c).
}
\end{figure*}

\begin{figure*}[htb]
\includegraphics[width=\columnwidth]{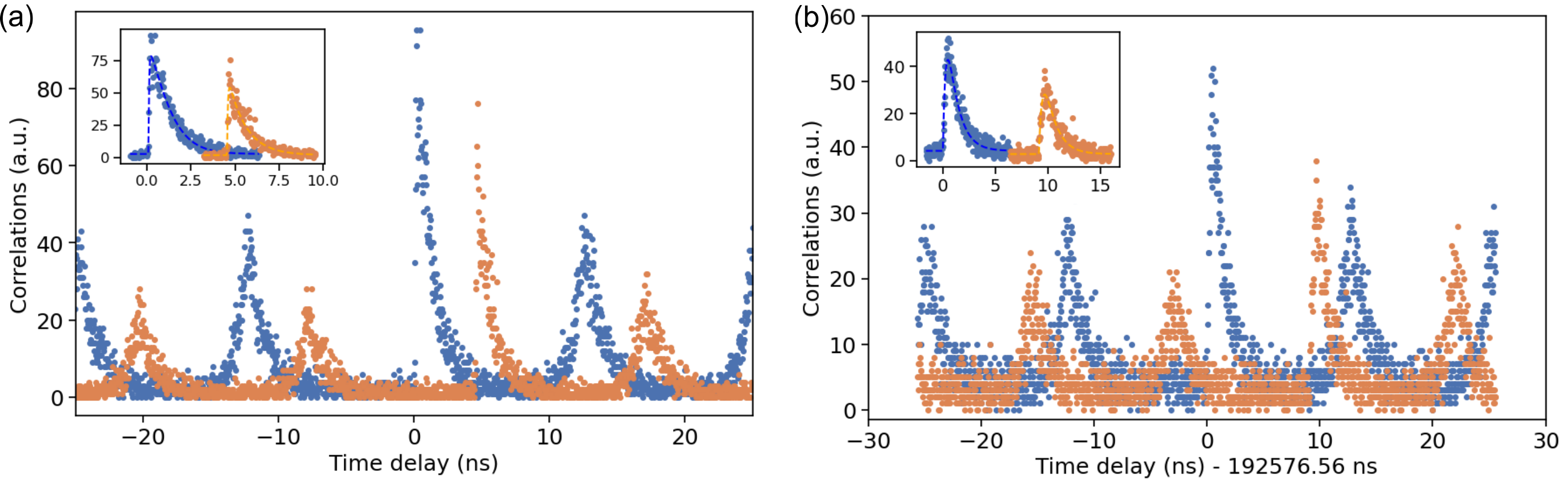}
\caption{\label{smfig:fig6-synch_QD2} 
\textbf{Clock synchronization.} (a) Correlations between the biexciton photons and the exciton photons measured between the two nodes, before (blue) and after (orange) adding the \SI{4.484}{\nano\second} fiber delay.
(b) Correlations between the biexciton photons and the returned exciton photons, both measured at the provider node, before (blue) and after (orange) the \SI{4.484}{\nano\second} fiber delay.
}
\end{figure*}
To further validate the presented entanglement-verified time synchronization scheme, we performed the same set of measurements using a different quantum dot source over the time-frame of a year. 
By comparing results across different quantum dots, we assess the robustness and reproducibility of the synchronization protocol and entanglement properties, reinforcing its applicability for scalable quantum communication networks.
The emission spectra of the alternative quantum dot (QD2) is shown in Figure~\ref{smfig:fig4-spectra_QD2} with the exciton and biexciton highlighted. Figure~\ref{smfig:fig5-local_ent_QD2} shows the result from the local entanglement characterization in the provider location. The HH and HV basis are shown in Figure~\ref{smfig:fig5-local_ent_QD2} (a), and the time evolved fidelity and concurrence are shown in Figure~\ref{smfig:fig5-local_ent_QD2} (b) and (c), respectively. It should be noted that the fine structure splitting of this quantum dot is higher than the previous quantum dot, evident by the higher modulation frequency in the coincidence and fidelity plots. The maximum fidelity measured for the \SI{240}{\pico\second} time delay at the highlighted time bins in Figure~\ref{smfig:fig5-local_ent_QD2} (b) and (c) is \num{0.896 \pm 0.023} and the concurrence at the same time bin is \num{0.832 \pm 0.041}.

For the entanglement distribution across the polarization-stabilized link connecting the provider and the subscriber, we measure a maximum fidelity of \num{0.711 \pm 0.133} and a maximum concurrence of \num{0.657 \pm 0.184}.

We performed the same time synchronization measurement discussed in the main text with and without having the \SI{4.484}{\nano\second} fiber delay. 
Figure~\ref{smfig:fig6-synch_QD2} (a) shows the cross-correlations between the biexciton photons and the transmitted exciton photons, measured at the subscriber node, before (blue) and after (orange) adding the fiber delay, revealing an absolute clock offset of \num{16471.248145206522}~seconds. 
Figure~\ref{smfig:fig6-synch_QD2} (b) shows the cross-correlation measured at the provider location between the biexciton photons and the returned exciton photons, before (blue) and after (orange) the fiber delay. Here we measure a round-trip time of \SI{192.576560}{\micro\second}, corresponding to an optical and electrical path difference of \SI{19663.828}{meter} between the two clocks. The difference between this value and the round-trip reported in the main text is due to the difference in the path length (optical and electrical) used in the two experiments.

After introducing the \SI{4.484}{\nano\second} fiber delay, the zero delay cross-correlations peak shift accordingly. From this measurement, we determine an added time delay of \SI{4.608}{\nano\second} due to the fiber, with a resulting shifted clock offset of \num{16471.248145210935}~seconds. These results confirm that the entanglement-assisted time synchronization method remains robust and consistent across different quantum dot sources with \num{10}s of ps precision.

\bibliography{sn-bibliography-SM}